\documentclass[11pt]{article}
\usepackage{times}
\usepackage{moreverb,amsmath}
\usepackage[margin=2.5cm]{geometry}

\usepackage{authblk}

\title{Simulating recurrent events that mimic actual data: a review of the literature with emphasis on event-dependence}
\author[a]{Juliette P\'{e}nichoux}
\author[a]{Thierry Moreau}
\affil[a]{Inserm U1018, CESP Centre for Research in Epidemiology and Population Health, Villejuif, France}
\author[b]{Aur\'{e}lien Latouche \thanks{aurelien.latouche@cnam.fr}}
\affil[b]{Conservatoire national des arts et m\'etiers, EA4629, Paris, France}

\date{}

\begin{document}
\maketitle

\begin{abstract}
We conduct a review to assess how the simulation of repeated or recurrent events are planned. For such multivariate time-to-events, it is well established that the underlying mechanism is likely to be complex and to involve in particular both heterogeneity in the population and event-dependence. In this respect, we particularly focused on these two dimensions of events dynamic when mimicking actual data. Next, we investigate whether the processes generated in the simulation studies have similar properties to those expected in the  clinical data of interest. Finally  we  describe a simulation scheme for generating data according to the time-scale of choice (gap time/ calendar) and to whether heterogeneity and/or event-dependence are to be considered. The main finding is that event-dependence is less widely considered in simulation studies than heterogeneity. This is unfortunate since  the occurrence of an event may alter the risk of occurrence of new events.
\end{abstract}


\maketitle

\section{Introduction}

Simulation studies that are performed for assessing statistical methods used in recurrent events analysis often rely on standardized event generation processes that cannot take into account some features of the data \cite{pmid16217859}. Indeed, it is well recognized in the literature on recurrent events that in most medical research applications, the underlying process is likely to be complex, involving both heterogeneity in the population, and event-dependence \cite{pmid16345026}. Heterogeneity arises if some subjects are more prone than others to experience events, owing to unkown or unmeasured factors, e.g. associated with lifestyle or genetic predisposition. This leads to intra-subject correlations between the waiting times separating consecutive events. Event-dependence refers to the case when the occurrence of one event may render the occurrence of subsequent ones more or less likely, e.g. as a result of a learning or immunity process, of a weakening of the body after an event or of an intervention subsequent to each event. The aim of this paper is to review the simulation methods that are used in the literature, focusing on those which generate data according to the features of actual data that are encountered in practice, so that they could be recommended for wider use. While, in many articles, heterogeneity is included in the model generating the simulated data, such is not the case of event-dependence, as will be seen. This work aims to describe the useful generating methods in detail and to underline the relevance of each of them for considering data of different types.   In section \ref{sec:literature}, the literature is reviewed in order to assess how recurrent events processes are generated, and how they match the events dynamic of the biomedical data motivating these articles. The simulations performed are compared with the clinical studies considered. Section \ref{meth.sim} describes methods for generating data according to the scenario of interest. A discussion follows in which the benefits of the methods are compared and some guidelines are suggested. 

\section{Results of the literature search}\label{sec:literature}

\subsection{Introduction to the intensity models used}

The simulation studies included in the following review mostly rely on modeling intensity or include processes that can be written as intensity-based processes. We focus on multiplicative proportional hazards intensities as they are the most frequent in the literature.  
The intensity process may be defined in one of two common timescales: the calendar-time and the gap-time. In calendar-time, the time denotes the time elapsed since the origin. In gap-time, the time is reset to zero when an event occurs, so that the time corresponds to the time elapsed since the previous event \cite{pmid10623910}.

In calendar-time, for a vector of external covariates $X$, the common multiplicative proportional intensity $\lambda(.)$ can be written:
\begin{equation}\label{agmod}
\lambda(t|X=x)=\lambda_0(t)\exp(\beta' x).
\end{equation}

In gap-time, the common multiplicative proportional intensity becomes:
\begin{equation}\label{renew.mod}
\lambda(t|X=x)=h_0(t-T_{N(t^-)})\exp(\beta' x).
\end{equation}

The models are described in more detail in section \ref{meth.sim}.

\subsection{How recurrent events processes are generated in the literature}

We conducted a literature research in six journals: Biometrics, Biometrika, Computational Statistics and Data Analysis, Journal of the Americal Statistical Association, Lifetime Data Analysis and Statistics in Medicine. This search was performed using the journals' webpages on October $10^{th}$, 2011, focusing on articles published between 1996 and this date. In total, we found 135 articles that featured the term 'recurrent events', 'repeated events' or 'repeated failures', in the
title and/or in the key words. Among these, 102 featured simulation studies of recurrent events. 

The generating intensity processes could be classified according to timescale, consideration of heterogeneity, and consideration of event-dependence. In the simple case where the baseline intensity is constant, the process can be expressed equivalently in either timescale (see Section \ref{meth.sim}), so Tables \ref{tablebiblio} and \ref{tablebiblio2} include a category "constant baseline" in addition to the categories "gap-time" and "calendar-time". A total of 92 articles could be classified in this way. In the sequel, the focus will be on these 92 articles.  The 10 remaining articles were not classified, for example because the focus was not on time of events but only on the number of events over given intervals. 

The results are summarized in Table \ref{tablebiblio} in which simulation studies are classified according to timescale and type of population (homogeneous or heteregeneous), and in Table \ref{tablebiblio2}, in which simulation studies are classified according to timescale and whether or not the generation processes included event-dependence. Note that one article can appear in more than one cell in each of these tables if several scenarios are considered in the simulation study. 

Among the $92$ articles, $54$ ($59\%$) featured exlusively processes with constant baseline. Note that a constant baseline does not exclude time-dependent covariates. Only $13$ $(14\%)$ considered processes with a baseline that was a function of time since the previous event (gap-time), and $28$ ($30\%$) featured processes with a non-constant calendar baseline. 

Among the articles with a baseline that varied with time, $3$ articles \cite{pmid11677832,pmid16320269,pmid17542002} also considered a timescale that was in-between the gap-time and the calendar-time: some events caused a "perfect repair" that led the baseline to restart at $0$, while for other events the baseline continued as in a calendar timescale.  

Heterogeneity among subjects was widely considered: $80$ articles ($87\%$) included in the generation process a subject-specific random effect that was usually gamma-distributed.

Event-dependence was more rarely considered : only $14\ (15\%)$ of the simulation studies feature this scenario. It was modeled through an internal covariate, an event-specific baseline or an event-specific random effect.

In addition, Table \ref{tablebiblio} suggests that the situation of a heterogeneous population is seldom considered when the baseline hazard is time-varying and formulated in gap-time (in 7 articles among the 13 featuring processes generated according to this timescale). Table \ref{tablebiblio2} suggests that the situation of event-dependence is less widely considered in simulation studies using a time-varying baseline hazard in a calendar timescale (in 2 articles \cite{pmid16320269,pmid18381708} among the 28 using this timescale).

The list of papers included in this review are available as supplementary material. 

\subsection{Comparison of generating process and data of interest in two examples}

In most of the simulation studies considered, the statistical methods investigated were also used in the analysis of biomedical data. In this section, we focus on two examples of applications that are commonly encountered in the recurrent events analysis setting, and investigate whether the processes generated in the simulation study have similar properties to those expected in the data of interest.

\subsubsection{Recurrent events associated with asthma}

Asthma is the disease of interest in several studies involving recurrent events. Three articles in our literature search analyze data from asthma studies \cite{pmid15293628,pmid18622700,pmid16135020}. The events of interest are not always the same. The focus was on hospitalisations or physicians' visits attributable to asthma in the study by Cai and Schaubel \cite{pmid15293628}, on exacerbations in the study by Chen and Cook \cite{pmid18622700}, and on coughing episodes in the study by Cook \emph{et~al}\cite{pmid16135020}. However, these different types of events are expected to reflect the general evolution of the disease. 

Asthma is expected to be a stable disease that does not evolve much with time, at least in adults \cite{asthma}, which is the population of interest in two of the studies mentioned \cite{pmid18622700,pmid16135020}. However, in young children, Cai and Schaubel \cite{pmid15293628} found that the risk of event was increased strongly between 0 and 2 years old and was decreased after 3 years old. Heterogeneity is to be expected as it is well-known that there are various degrees of severity of asthma \cite{Wildfire2012}. The fact that asthma is considered as a stable disease suggests that at least in adults there is no event-dependence and no time-dependence.

The three simulation studies considered \cite{pmid15293628,pmid18622700,pmid16135020} all relied on simulated processes that had a constant baseline and featured no event-dependence but were heterogeneous across the sample. This matches the expected events dynamic of the data processes, except maybe for the asthma study by Cai and Schaubel \cite{pmid15293628} which focuses on children, for whom time-dependence is to be expected.

\subsubsection{Recurrent infections in patients with chronic granulomatous disease}

Among the articles considered in the review of simulation studies, seven \cite{pmid16345026,Lin2000,pmid18516715,pmid20390350,pmid21590791,pmid19034646,Zeng2007a} investigated an application to a dataset from a clinical trial that was originally introduced by Flemming and Harrington \cite{Fleming2005},  which studies the effect of gamma interferon in patients with chronic granulomatous disease (CGD). This double-blind trial focused on the occurrence of recurrent infections among 128 patients randomised between placebo and gamma interferon. 

The expected clinical course of CGD is not well-known since it is a relatively rare condition (about one in 250000 individuals) \cite{Berg2009}. It is therefore not obvious whether one should expect time-varying hazards, but in the absence of evidence this possibility should not be discarded. Moreover, one study \cite{pmid18037347} pointed out that the incidence for some particular types of infections increases during follow-up.  Similarly, event-dependence has received little attention, but the analysis of the dataset of the gamma interferon trial by Box-Steffensmeier and De Boef \cite{pmid16345026} provides an indication of event dependence, specifically that the occurrence of an infection makes recurrence more likely. Additionally, it is well-known that there is much clinical heterogeneity among patients with CGD \cite{Dinauer2005,pmid10844935}.

Among the seven simulation studies presented in articles featuring an application to the CGD data, two considered only homogeneous processes \cite{pmid20390350,pmid21590791}. A time-varying baseline in a calendar timescale was considered in four \cite{pmid20390350,pmid18516715,pmid19034646,Zeng2007a} of the seven studies, which is more than expected given the total proportion found in the review (30$\%$). Event-dependence, however, was considered only in one simulation study \cite{pmid16345026}.

\section{Simulations based on intensity-based models}\label{meth.sim}

The simulation studies included in the review in Section \ref{sec:literature} were based on the modeling intensity or could be written as intensity-based processes. This intensity could be additive or more commonly multiplicative. Accelerated failure time models, which correspond to non-proportional intensities, are seldom used in this setting. In this section, we present examples of intensity processes for each scenario defined in Section \ref{sec:literature}. We focus on multiplicative proportional hazards intensities as they are the most frequent in the literature.  

For a given suject, let $T_j$ denote the time elapsed from the origin to the $j^{th}$ event, $T_0=0$, and let $W_j=T_j-T_{j-1}$ denote the $j^{th}$ waiting time between two consecutive events. The $W_j$ are referred to as gap times.

Whichever intensity process is considered, a first step is to generate a censoring time C, and external covariates X if necessary. Then, the event times $T_j$ (or the gap times $W_j$) are generated until the censoring time is reached. The censoring time is usually generated according to an exponential or uniform distribution.

\subsection{Standard model in each timescale}

If the baseline is constant, the recurrent event process can be considered equivalently as an homogeneous Poisson process or a sequence of exponentially distributed gaps, which makes the generation process straightforward. A constant baseline is the most widely considered scenario. However, it may be beneficial to consider different generation processes that involve time-varying baselines using the two possible timescales.

\subsubsection{Calendar timescale}\label{sec:ct}

When the relevant timescale is the time elapsed since inclusion, the baseline is specified in a calendar timescale. In particular, calendar-time is the proper timescale in the case of a disease that evolves in the long run. For example, the calendar timescale is used when analyzing exacerbations in patients with chronic bronchitis, which often arise as a consequence of impaired lung function due to the underlying disease \cite[p233]{recevents}.

The standard model is a non-homogeneous Poisson process, with an intensity given by (\ref{agmod}). 

The most common choice is a Weibull intensity, with $\lambda_0(t)=\lambda\nu t^{\nu-1}$ where $\lambda$ is a scale parameter and $\nu$ a shape parameter. The Weibull distribution allows the baseline to increase (if $\nu>1$) or decrease (if $\nu<1$) over time.

There are several ways to simulate a non-homogeneous Poisson process. 

The first is the inversion method. The $j^{th}$ gap time $W_j$ of a non-homogeneous Poisson process has a cumulative distribution function $F_j(.)$ equal to:

\begin{align}
F_j(w) &=1-P(W_j > w)\nonumber\\
&=1-P\{N(T_{j-1}+w)-N(T_{j-1})=0\}.\label{poisson.incr}
\end{align}

The number of events observed for a non-homogeneous Poisson process with intensity $\lambda(.)$ between two times $s$ and $t$ follows a Poisson distribution with parameter $\int_s^t\lambda(u)du$, which leads to:

\begin{equation}\label{proba}
N(T_{j-1}+w)-N(T_{j-1})\sim \mathcal{P}\left(\int_{T_{j-1}}^{T_{j-1}+w}\lambda(u)du\right).
\end{equation}

And (\ref{poisson.incr}) becomes :

\begin{equation}\label{sim.calend.meth1}
F_j(w)=1-\exp\left(-\int_{T_{j-1}}^{T_{j-1}+w}\lambda(u)du\right).
\end{equation}

The successive event times are then simulated according to the following algorithm:
\begin{enumerate}
	\item Let $T_0=0$.
	\item For the $j^{th}$ event, simulate $V_j\sim\mathcal{U}[0,1]$.
	\item Let $W_j=F_j^{-1}(V_j)$ and $T_j=T_{j-1}+W_j$.
	\item If $T_j<C$, return to step 2 for the next event. Otherwise, the $j^{th}$ event time is censored: set $T_j=C$ and $W_j=C-T_{j-1}$.
\end{enumerate}

In the Weibull intensity in model (\ref{agmod}), the event times simulated are given by:

\begin{equation*}
T_j=T_{j-1}+W_j=\left(-\frac{\log(1-V_j)}{\lambda\exp(\beta'x)}+T_{j-1}^\nu\right)^{\frac{1}{\nu}}.
\end{equation*}

Note that if $V_j\sim\mathcal{U}[0,1]$, then $1-V_j\sim\mathcal{U}[0,1]$, and $\log(1-V_j)$ can be replaced by $\log(V_j)$ in this expression.

A limitation of this method is that it requires $F_j(.)$ to be invertible. If not, numerical methods may be used to obtain $W_j$ such that $F_j(W_j)=V_j$.

Another common method is referred to as thinning \cite{Lewis1979}. First, a value $\overline\lambda$ is chosen such that $\forall t, \lambda(t)\leq \overline\lambda$. If  times $T_1^*, T_2^*...$ are generated from a homogeneous Poisson process with rate $\overline\lambda$, the process in which each time $T_j^*$ has a probability $\frac{\lambda(T_j^*)}{\overline\lambda}$ of being an actual event time, and the complementary probability $1-\frac{\lambda(T_j^*)}{\overline\lambda}$ to be discarded, is a process with intensity $\lambda(.)$. The corresponding generating algorithm is as follows:
\begin{enumerate}
	\item Let $T_0=0$, $T^*=0$, $j=1$.
	\item Simulate $E\sim Exp(\overline\lambda)$, and let $T^*=T^*+E$.
	\item Simulate $V\sim\mathcal{U}[0,1]$: if $V\leq\frac{\lambda(T_j^*)}{\overline\lambda}$, $T^*=T_j$ is the $j^{th}$ event time. Then set $j=j+1$, and return to step 2. If $V>\frac{\lambda(T_j^*)}{\overline\lambda}$, $T^*$ is not an event time, go back directly to step 2.
\end{enumerate}

The assumption that the intensity $\lambda(.)$ is an  upper-bounded by a value $\overline\lambda$ is not as restrictive as it might seem, since processes are usually generated over a finite time interval \cite{recevents}.
This second method is often slower than the inversion method and is usually used when the inverse of the cumulative distribution functions $F_j(.)$ is not easily calculated. 

\subsubsection{Gap timescale}\label{sec:gt}

When the relevant timescale is the time elapsed since the previous event, the baseline is specified in gap-time. The gap-time is relevant for example when generating a risk of relapse after a complete response to treatment \cite{pmid16320269}. The basic model is a renewal process, with an intensity given by (\ref{renew.mod}).

The Weibull distribution is again a natural choice, with $h_0(w)=\lambda\nu w^{\nu-1}$ allowing for a risk of event that increases or decreases starting from the occurrence of the previous event. 

The successive i.i.d. gap times can be simulated in the same way survival times are simulated when the hazard function is time-varying \cite{pmid15724232}. The cumulative distribution function for the $j^{th}$ gap time $W_j$ is : $F_j(w)=1-\exp\{H_0(w)\exp(\beta'x)\}$. 
Since $F_j(W_j)\sim\mathcal{U}[0,1]$, $\exp\{-H_0(w)\exp(\beta'x)\}\sim\mathcal{U}[0,1]$. The gap time $W_j$ can therefore be simulated by generating $V_j\sim\mathcal{U}[0,1]$, and setting $W_j=H_0^{-1}\left[-\frac{\log(V_j)}{\exp(\beta'x)}\right]$. In the case of the Weibull distribution, it leads to $W_j=\left(-\frac{\log(V_j)}{\lambda\exp(\beta'x)}\right)^{\frac{1}{\nu}}$.

New gap times are simulated until the sum of the gap times exceeds the censoring time: the last gap time is then censored. 

When the cumulative hazard is not easily invertible, one may use numerical methods to solve $H_0(W_j)\exp(\beta'x)\approx -\log(V_j)$.

Again, an alternative method is to use an acceptance-rejection type algorithm \cite[p170]{Leemis1990} adapted from the thinning method proposed by Lewis and Shedler  for non-homogeneous Poisson processes \cite{Lewis1979}. 

Denote $\overline h$ an upper-bound of the hazard function h(.) over the period considered. Each gap time $W_j$ is simulated according to the following algorithm:
\begin{enumerate}
	\item Set $W_0=0$ and $W^*=0$.
	\item Generate $E\sim Exp(\overline h)$, and let $W^*=W^*+E$.
	\item Generate $V_j\sim\mathcal{U}[0,1]$. If $V_j\leq\frac{h(W^*)}{\overline h}$, $W^*$ is the $j^{th}$ simulated gap time $W_j$. Otherwise return to step 2. 
\end{enumerate}
Again, new independent gap times are simulated until the sum of the gap times exceeds the censoring time: the last gap time is then censored. 

\subsection{Heterogeneity}

Heterogeneity in the population is often realistic in biomedical studies and is widely considered in simulation studies designed to evaluate the properties of statistical methods. In both timescales, heterogeneity can be easily introduced by generating a subject-specific random effect. This random effect is usually assumed to follow gamma or lognormal distribution, but other distributions are possible. For instance, one may use a binary random effect to investigate model properties when binary covariates are omitted.

In the calendar timescale, the resulting process is referred to as a mixed Poisson process, i.e. the individual processes are Poisson processes with subject-specific intensities. The usual multiplicative proportional intensity can then be written :
\begin{equation}\label{ajnum}
\lambda(t|U=u,X=x)=\lambda_0(t)u\exp(\beta' x),
\end{equation}

where $u$ is the subject-specific random effect. A simple simulation scheme consists in first generating independent individual frailties according to the chosen distribution. The individual intensities are obtained from (\ref{ajnum}). The individual processes can then be simulated as Poisson processes as described in Section \ref{sec:ct}.

In gap-time, the result is a shared frailty model, with the corresponding intensity :
\[
\lambda(t|U=u,X=x)=h_0(t-T_{N(t^-)})u\exp(\beta' x).
\]
Similarly, independent individual frailties are first generated. The gap times for each subject can then be simulated as described in Section \ref{sec:gt}.
 
\subsection{Generating recurrent event with event-dependence }

Event-dependence is less widely considered in simulation studies than heterogeneity. However, for many recurrent events data, it is realistic to consider that the occurrence of an event may alter the risk of occurrence of new events. For this reason, it is useful to investigate the properties of statistical methods under event-dependence, e.g. when the occurrence of an event increases the risk of occurrence of subsequent ones. 

\subsubsection{Event-dependence in gap-time}

It should be noted that a baseline that is a function of time in gap-time constitutes a form of event-dependence, since the occurrence of an event causes the hazard to reset back to its original level. However, additional event-dependence may be appropriate. A common assumption is to consider that the occurrence of an event increases the risk of further ones (e.g. by making subjects more frail and therefore prone to experiencing events), or on the contrary decreases it (by making subjects immune or more vigilant). A natural way to generate such data is to take an event-specific baseline or an internal time-dependent covariate that induces dependence \cite{kalbfleisch}, as did Cook \emph{et~al} \cite{Cook2009a} in their simulation studies.

A simple example is to multiply the baseline by $\alpha^{j-1}$ with $j-1$ the number of previous events. Setting $\alpha<1$ will cause the risk of event to decrease with the number of past events while setting $\alpha>1$ will cause the risk to increase with the number of events. This is similar to including the number of previous events as an internal covariate in a Cox-type model with associated parameter $\ln(\alpha)$.

However, if the risk increases exponentially after each event, the event process may "explode", with a number of events that tends to infinity for a finite time \cite{pmid20625827}. A common solution is to consider the risk to be stabilised after a pre-specified number of events, e.g. 4 events. Considering a Weibull baseline and a risk that is stable after 4 events, the data generation model without covariates becomes :

\begin{equation*}
h_{0j}(w) = 
\begin{cases}
 \lambda\nu w^{\nu-1}\alpha^{j-1} &j=1,2,3,4 \\
 \lambda\nu w^{\nu-1}\alpha^4 &j>4. \\
\end{cases}
\end{equation*}

Usually, when considering a gap-timescale, the internal covariate that induces the dependence between events is constant for each gap-time and includes the number or timing of previous events. The successive event times $W_j$ can then be simulated as described in Section \ref{sec:gt}, with a different cumulative distribution function for each event.

However, more general processes can be considered, with internal time-dependent covariates that can be non-constant over the at-risk interval for a given event. The methods described in the next section for calendar-time are general and valid for such processes.

\subsubsection{Event dependence in calendar-time: generating dynamic covariates}

Inducing event-dependence in a calendar timescale is usually done by considering an intensity with an internal (or dynamic) covariate that depends on the past of the process. An obvious choice is $N(t^-)$ to allow the intensity to increase or decrease with the number of past events.  Such a generating process was considered by Gonzalez \emph{et~al} \cite{pmid16320269}. However, just as in gap-time, the risk is that the number of events will increase enormously when considering $N(t^-)$ as a covariate. Other possible covariates are $N(t^-)/t$, in order to simulate processes in which each event leads to an increased intensity, but where the effect of this increase fades with time, $min(N(t^-),K_0)$ to consider the case where the risk becomes stable after $K_0$ events, or $\frac{N(t^-)-N(t-u)}{u}$ where $u\leq t$, in order to focus on recent history (on a period of length $u$) as suggested by Elgmati \emph{et~al} \cite{pmid19701791}. 

The inversion method described in Section \ref{sec:ct} for non-homogeneous Poisson processes can be used for more general processes, with an intensity that depends on the past. Indeed, Cook and Lawless \cite[p30]{recevents} showed that:

\begin{equation*}
P(W_j>w|T_{j-1}=t_{j-1},H(t_{j-1}))=\exp\left(-\int_{t_{j-1}}^{t_{j-1}+w}\lambda(u|H(u))du\right),
\end{equation*}
where H(t) is the history of the event process up to time $t$ (not included). The intensity is therefore considered deterministic on the interval $]t_{j-1},t_{j-1}+w]$. This makes it possible to simulate a process defined by an intensity conditional on the past by the inversion method in the same way as a non-homogeneous Poisson process, by generating $F_j(W_j)\sim\mathcal{U}[0,1]$ with $F_j(w)=P(W_j\leq w|T_{j-1},H(T_{j-1}))$. 

For example, the process with intensity:

\begin{equation*}
\lambda(t)=\lambda_0(t)\exp(\beta'x+\varphi N(t^-)),
\end{equation*}

has a cumulative distribution function for the $j^{th}$ gap time:

\begin{equation*}
F_j(w)=1-\exp\left(-\int_{t_{j-1}}^{t_{j-1}+w}\lambda_0(u)\exp(\beta'x+\varphi(j-1))du\right).
\end{equation*}

When the baseline intensity is Weibull, generating $1-F_j(W_j)=V_j\sim\mathcal{U}[0,1]$ leads to:

\begin{equation*}
T_j=T_{j-1}+W_j=\left(-\frac{\log(V_j)}{\lambda\exp(\beta'x+\varphi(j-1))}+T_{j-1}^\nu\right).
\end{equation*}

When considering the covariate $min(N(t^-),K_0)$ with a Weibull baseline intensity, the event times become:

\begin{equation*}
T_j=\left(-\frac{\log(V_j)}{\lambda\exp(\beta'x+\varphi(min(j-1,K_0)))}+T_{j-1}^\nu\right)^{\frac{1}{\nu}}.
\end{equation*}

More generally, an intensity process:

\begin{equation*}
\lambda(t)=\exp(g_0(t)+I(N(t^-)>0)g_1(t- N(t^-))+g_2(N(t^-))+\beta'x),
\end{equation*}
 which is a function of the time in a calendar timescale, the time in gap-time and the number of past events, leads to a cumulative distribution function of the $j^{th}$ gap time :

\begin{equation}\label{general.intensity}
F_j(w)=\exp\left(-\int_{0}^{w}\exp\{g_0(t_{j-1}+u)+I(j-1>0)g_1(u)+g_2(j-1)+\beta'x\}du\right).
\end{equation} 
  
$F_j(W_j)\sim\mathcal{U}[0,1]$ is then generated and $W_j$ is obtained from (\ref{general.intensity}) analytically or numerically. In the case of the internal covariates $N(t^-)/t$ or $\frac{N(t^-)-N(t-u)}{u}$ suggested before, a closed-form expression for $W_j$ cannot easily be obtained.

Another possibility is to use the thinning method described in Section \ref{sec:ct}, which was introduced by Lewis and Shedler \cite{Lewis1979} for non-homogeneous Poisson processes, but was later proved to be valid when the intensity depends on the past \cite{Ogata1981}. In the algorithm described in Section \ref{sec:ct}, $\lambda(T^*)$ can therefore depend on the past.

A final possibility that does not require an upper bound for the intensity or a numeric inversion of the cumulative distribution function is based on a discrete approximation. A similar method was used for example by \cite{Aalen2008}. A partition of the time period is defined by intervals of length $\Delta t$: $[k\Delta t,(k+1)\Delta t]$, $k=0,1,2,...$. The length $\Delta t$ is chosen to be small, so that one can assume that a subject cannot have more than one event in each interval $[k\Delta t,(k+1)\Delta t]$. The occurrence of an event during the interval $[k\Delta t,(k+1)\Delta t]$ is generated with an event probability $\lambda(k\Delta t)\Delta t$, which approximates the probability of an event occurrence corresponding to (\ref{proba}). The intensity is then updated for the next interval. For example, if $\Delta t$ corresponds to a day, the occurrence or not of an event is simulated every day. According to this approximation, simulating a process with an intensity  $\lambda_0(t)\exp(\beta'x+\varphi\frac{N_i(t^-)}{t})$ where $\lambda_0(.)$ is expressed as a function of time in days leads to consider that the subject has an event on the $k^{th}$ day with a probability $\lambda_0(k)\exp(\beta'x+\varphi\frac{N_i(k-1)}{k-1})$. This method of simulating general event processes is simple, but can also be slow. 

\section{Discussion}

This review complements the extensive simulation study conducted by Metcalfe and Thompson \cite{pmid16217859}. Their study exemplified the fact that no model is adequate for all settings, and that it is necessary to investigate several situations in simulation studies, including heterogeneity and event-dependence, in order to have satisfactory knowledge of the properties of a given statistical method.

In this review, we show that heterogeneity has received much attention. By contrast, event-dependence is seldom considered in simulation studies, especially when the event-generation process is expressed in calendar-time. This is unfortunate since event-dependence is known to have a critical effect on statistical models, in terms of properties and interpretation of the covariates effects \cite{Cheung2010}. One reason for neglecting this possibility may be that the generation of non-Poisson processes is not obvious. In this paper, the general relation between the intensity process and the successive gap times is recalled and examples of event-dependent processes are provided, in both timescales. Additionally, as noted by Bender \emph{et~al} \cite{pmid15724232} for univariate survival times, most simulation studies use only constant baseline hazards corresponding to exponential gap times. However, it was pointed out by Duchateau \emph{et~al} \cite{asthma} that the choice of the timescale in the statistical model may considerably impact the results. This suggests that a given model may demonstrate different properties in simulation studies according to the timescale of the generation process, and that it would be wise to consider generation processes with time-varying baselines in both timescales. 

Extensive simulation studies that investigate the properties of a given statistical method should therefore include, in our opinion, datasets generated by using a mixed Poisson process, processes with constant baseline, Weibull calendar-time, and Weibull gap-time baseline, and processes with event-dependence e.g. with the number of previous events included as a covariate in the intensity generation process. Such studies would assist the applied statistician in choosing the appropriate method in a particular biomedical context.
When a statistical method is suggested for a biomedical context, the scenario considered for the simulation studies should match the expected dynamics of the event-process in this context. If a long-term evolution of the disease cannot be ruled out, calendar-time-varying intensity processes should be included. This would be the case for exacerbations in patients with chronic brochitis \cite[p233]{recevents}. A baseline varying with time in gap-time should be considered in the case when an event is followed by an intervention that leads to a kind of \textit{renewal} of the subject. This was considered in particular in the context of cancer relapse where a complete response was possible \cite{pmid16320269}. Heterogeneity should in general be included since there is no biomedical context in which one can be sure that all risk factors are known and always measured. Event-dependence should be considered in contexts in which an event could alter the risk of subsequent events. For example, this was considered in a simulation study by Kvist \emph{et~al} \cite{pmid18381708}, in the context of affective disorder, in which an important theory is the sensitisation theory, i.e. that the occurrence of an event may increase the risk of subsequent events. 

\pagebreak

\newpage
\begin{table}%
\caption{Results of literature search : generated processes classified according to timescale and type of population (homogeneous/heterogeneous) (cells are non-exclusive)}
\centering
\footnotesize
\begin{tabular}{lcc|c}
\hline
&Homogeneity	&Heterogeneity  &Total\\
	\hline
Constant baseline hazard &28 (30$\%$)&62 (67$\%$)&70 (76$\%$)\\
Baseline hazard in gap-time &8 (9$\%$) &7 (8$\%$)&13 (14$\%$)\\  
Baseline hazard in calendar-time &12 (13$\%$)&25 (27$\%$) &28 (30$\%$) \\  
\hline
	Total	&38 (41$\%$)&80 (87$\%$)&92\\
	\hline
\end{tabular}
\label{tablebiblio}
\end{table}

\begin{table}%
\caption{Results of literature search : generated processes classified according to timescale and inclusion of event-dependence (cells are non-exclusive)}
\centering
\footnotesize
\begin{tabular}{lcc|c}
\hline
&No event-dependence	&Event-dependence  &Total\\
	\hline
Constant baseline hazard &64 (70$\%$) &13 (14$\%$) &70 (76$\%$)\\
Baseline hazard in gap-time &12 (13$\%$) &3 (3$\%$) &13 (14$\%$)\\  
Baseline hazard in calendar-time  &27 (30$\%$) &2 (2$\%$) &28 (30$\%$)\\  
\hline
	Total	&83 (90$\%$) &14 (15$\%$) &92\\
	\hline
\end{tabular}
\label{tablebiblio2}
\end{table}


\section{Bibliography}

\begin{thebibliography}{99}
\bibitem{pmid16217859}
Metcalfe C, Thompson SG. {{T}he importance of varying the event generation
  process in simulation studies of statistical methods for recurrent events}.
  \emph{Statistics in Medicine}  Jan 2006; \textbf{25}:165--179.

\bibitem{pmid16345026}
Box-Steffensmeier J, De~Boef S. {{R}epeated events survival models: the
  conditional frailty model}. \emph{Statistics in Medicine}  Oct 2006;
  \textbf{25}:3518--3533.

\bibitem{pmid10623910}
Kelly P, Lim L. {{S}urvival analysis for recurrent event data: an application
  to childhood infectious diseases}. \emph{Statistics in Medicine}  Jan 2000;
  \textbf{19}:13--33.

\bibitem{pmid11677832}
Agustin MZ, Pena EA. {{G}oodness-of-fit of the distribution of
  time-to-first-occurrence in recurrent event models}. \emph{Lifetime Data
  Analysis}  Sep 2001; \textbf{7}:289--306.

\bibitem{pmid16320269}
Gonzalez JR, Pena EA, Slate EH. {{M}odelling intervention effects after cancer
  relapses}. \emph{Statistics in Medicine}  Dec 2005; \textbf{24}:3959--3975.

\bibitem{pmid17542002}
Han J, Slate EH, Pena EA. {{P}arametric latent class joint model for a
  longitudinal biomarker and recurrent events}. \emph{Statistics in Medicine}
  Dec 2007; \textbf{26}:5285--5302.

\bibitem{pmid18381708}
Kvist K, Andersen PK, Angst J, Kessing LV. {{R}epeated events and total time on
  test}. \emph{Statistics in Medicine}  Aug 2008; \textbf{27}:3817--3832.

\bibitem{pmid15293628}
Cai J, Schaubel DE. {{M}arginal means/rates models for multiple type recurrent
  event data}. \emph{Lifetime Data Analysis}  Jun 2004; \textbf{10}:121--138.

\bibitem{pmid18622700}
Chen BE, Cook RJ. {{T}he analysis of multivariate recurrent events with
  partially missing event types}. \emph{Lifetime Data Analysis}  Mar 2009;
  \textbf{15}:41--58.

\bibitem{pmid16135020}
Cook RJ, Wei W, Yi GY. {{R}obust tests for treatment effects based on censored
  recurrent event data observed over multiple periods}. \emph{Biometrics}  Sep
  2005; \textbf{61}:692--701.

\bibitem{asthma}
Duchateau L, Janssen P, Kezic I, Fortpied C. Evolution of recurrent asthma
  event rate over time in frailty models. \emph{Journal of the Royal
  Statistical Society. Series C, Applied Statistics}  2003;
  \textbf{52}:355--363.

\bibitem{Wildfire2012}
Wildfire J, Gergen P, Sorkness C, Mitchell H, Calatroni A, Kattan M, Szefler S,
  Teach S, Bloomberg G, Wood R, \emph{et~al.}. Development and validation of
  the composite asthma severity index-an outcome measure for use in children
  and adolescents. \emph{The Journal of Allergy and Clinical Immunology}  2012;
  \textbf{129}:694--701.

\bibitem{Lin2000}
Lin D, Wei L, Yang I, Ying Z. Semiparametric regression for the mean and rate
  functions of recurrent events. \emph{Journal of the Royal Statistical
  Society. Series B, Statistical Methodology}  2000; \textbf{62}:711--730.

\bibitem{pmid18516715}
Sun L, Su B. {{A} class of accelerated means regression models for recurrent
  event data}. \emph{Lifetime Data Analysis}  Sep 2008; \textbf{14}:357--375.

\bibitem{pmid20390350}
Sun L, Tong X, Zhou X. {{A} class of {B}ox-{C}ox transformation models for
  recurrent event data}. \emph{Lifetime Data Analysis}  Apr 2011;
  \textbf{17}:280--301.

\bibitem{pmid21590791}
Sun L, Zhou X, Guo S. {{M}arginal regression models with time-varying
  coefficients for recurrent event data}. \emph{Statistics in Medicine}  Aug
  2011; \textbf{30}:2265--2277.

\bibitem{pmid19034646}
Tong X, Zhu L, Sun J. {{V}ariable selection for recurrent event data via
  nonconcave penalized estimating function}. \emph{Lifetime Data Analysis}  Jun
  2009; \textbf{15}:197--215.

\bibitem{Zeng2007a}
Zeng D, Lin D. Semiparametric transformation models with random effects for
  recurrent events. \emph{Journal of the American Statistical Association}
  2007; \textbf{102}:167--180.

\bibitem{Fleming2005}
Fleming T, Harrington D. \emph{Counting processes and survival analysis}.
  Wiley, 1991.

\bibitem{Berg2009}
van~den Berg J, van Koppen E, Åhlin A, Belohradsky B, Bernatowska E, Corbeel L,
  Español T, Fischer A, Kurenko-Deptuch M, Mouy R, \emph{et~al.}. Chronic
  granulomatous disease: The european experience. \emph{PLoS ONE}  04 2009;
  \textbf{4}(4):e5234.

\bibitem{pmid18037347}
Martire B, Rondelli R, Soresina A, Pignata C, Broccoletti T, Finocchi A, Rossi
  P, Gattorno M, Rabusin M, Azzari C, \emph{et~al.}. {{C}linical features,
  long-term follow-up and outcome of a large cohort of patients with {C}hronic
  {G}ranulomatous {D}isease: an {I}talian multicenter study}. \emph{Clin.
  Immunol.}  Feb 2008; \textbf{126}(2):155--164.

\bibitem{Dinauer2005}
Dinauer M. Chronic granulomatous disease and other disorders of phagocyte
  function. \emph{American Society of Hematology Education Program Book}  2005;
  \textbf{2005}:89--95.

\bibitem{pmid10844935}
Winkelstein JA, Marino MC, Johnston RB, Boyle J, Curnutte J, Gallin JI, Malech
  HL, Holland SM, Ochs H, Quie P, \emph{et~al.}. {{C}hronic granulomatous
  disease. {R}eport on a national registry of 368 patients}. \emph{Medicine
  (Baltimore)}  May 2000; \textbf{79}(3):155--169.

\bibitem{recevents}
Cook R, Lawless J. \emph{The Statistical Analysis of Recurrent Events}.
  Springer, 2007.

\bibitem{Lewis1979}
Lewis P, Shedler G. Simulation of nonhomogenous poisson processes by thinning.
  \emph{Naval Research Logistics Quarterly}  1979; \textbf{26}:403--413.

\bibitem{pmid15724232}
Bender R, Augustin T, Blettner M. {{G}enerating survival times to simulate
  {C}ox proportional hazards models}. \emph{Statistics in Medicine}  Jun 2005;
  \textbf{24}:1713--1723.

\bibitem{Leemis1990}
Leemis L, Shih LH, Reynertson K. Variate generation for accelerated life and
  proportional hazards models with time dependent covariates. \emph{Statistics
  \& Probability Letters}  1990; \textbf{10}:335--339.

\bibitem{kalbfleisch}
Kalbfleisch JD, Prentice RL. \emph{The statistical analysis of failure time
  data}. Probability and statistics, Wiley, 2002.

\bibitem{Cook2009a}
Cook R, Lawless J, Lakhal-Chaieb L, Lee K. Robust estimation of mean function
  and treatment effects for recurrent events under event-dependent censoring
  and termination: Application to skeletal complications in cancer metastatic
  to bone. \emph{Journal of the American Statistical Association}  march 2009;
  \textbf{104}(485):60--75.

\bibitem{pmid20625827}
Gjessing HK, R?ysland K, Pena EA, Aalen OO. {{R}ecurrent events and the
  exploding {C}ox model}. \emph{Lifetime Data Analysis}  Oct 2010;
  \textbf{16}:525--546.

\bibitem{pmid19701791}
Elgmati E, Farewell D, Henderson R. {{A} martingale residual diagnostic for
  longitudinal and recurrent event data}. \emph{Lifetime Data Analysis}  Mar
  2010; \textbf{16}:118--135.

\bibitem{Ogata1981}
Ogata Y. On {L}ewis' simulation method for point processes. \emph{IEEE
  Transactions on Information Theory}  1981; \textbf{27}:23--31.

\bibitem{Aalen2008}
Aalen O, Borgan O, Gjessing H. \emph{Survival and event history analysis : a
  process point of view}. Springer, 2008.

\bibitem{Cheung2010}
Cheung Y, Xu Y, Tan S, Cutts F, Milligan P. {{E}stimation of interention
  effects using first or multiple episodes in clinical trials: the
  {A}ndersen-{G}ill model re-examined}. \emph{Statistics in Medicine}  2010;
  \textbf{29}:328--336.
	
\end{thebibliography}
\end{document}